\documentclass[final,5p,times,twocolumn]{elsarticle}

\usepackage{mathtools}
\usepackage{braket} 
\usepackage[dvipsnames]{xcolor}  
\usepackage{float}
\usepackage[T1]{fontenc}
\usepackage[utf8]{inputenc}
\usepackage{graphicx,color,rotating,pifont}
\usepackage{amsmath,amssymb}
\usepackage{mathrsfs}
\usepackage{mathtools}
\usepackage{subfigure}
\usepackage{siunitx}
\usepackage{bm}
\usepackage{ae}
\usepackage{dcolumn}
\usepackage{txfonts}
\usepackage{tensor}
\usepackage{braket}
\usepackage{booktabs}
\usepackage{xspace}
\usepackage{soul}
\usepackage{multirow}
\usepackage{tikz}
\usepackage{pgffor} 
\setstcolor{red}
\setul{}{.2ex} 
 
\usepackage[normalem]{ulem}

\usetikzlibrary{backgrounds}
\usetikzlibrary{shapes,matrix,trees}
\usetikzlibrary{arrows.meta}
\usetikzlibrary{positioning}			
\usetikzlibrary{calc,through}			
\usetikzlibrary{positioning,calc,shapes,decorations.pathreplacing,decorations.markings,decorations.pathmorphing}
\tikzset{
	vector/.style={decorate, decoration={snake,amplitude=2.5pt}, draw},
	provector/.style={decorate, decoration={snake,amplitude=2.5pt}, draw},
	antivector/.style={decorate, decoration={snake,amplitude=-2.5pt}, draw},
	fermion/.style={draw=black, postaction={decorate},
		decoration={markings,mark=at position .6 with {\arrow[draw=black]{>}}}},
           vL/.style={draw=ppurple, postaction={decorate},
		decoration={markings,mark=at position .6 with {\arrow[draw=ppurple]{>}}}},
	vLp/.style={draw=ppurple, postaction={decorate},
		decoration={markings,mark=at position .7 with {\arrow[draw=ppurple]{>}}}},
	NR/.style={draw=ggreen, postaction={decorate},
		decoration={markings,mark=at position .62 with {\arrow[draw=ggreen]{>}}}},	
	NRp/.style={draw=ggreen, postaction={decorate},
		decoration={markings,mark=at position .7 with {\arrow[draw=ggreen]{>}}}},	
	neutralino/.style={draw=black},
	fermionbar/.style={draw=black, postaction={decorate},
		decoration={markings,mark=at position .6 with {\arrow[draw=black]{<}}}},
	fermionnoarrow/.style={draw=black},
	gluon/.style={decorate, draw=black,
		decoration={coil,amplitude=4pt, segment length=5pt}},
	scalar/.style={dashed,draw=black, postaction={decorate},
		decoration={markings,mark=at position .55 with {\arrow[draw=black]{>}}}},
	scalarbar/.style={dashed,draw=black, postaction={decorate},
		decoration={markings,mark=at position .55 with {\arrow[draw=black]{<}}}},
	scalarnoarrow/.style={dashed,draw=black},
	electron/.style={draw=black, postaction={decorate},
		decoration={markings,mark=at position .55 with {\arrow[draw=black]{>}}}},
	bigvector/.style={decorate, decoration={snake,amplitude=4pt}, draw},
	photon/.style={decorate, draw=black,decoration={snake,amplitude=4pt, segment length=5pt} }
}

\usepackage[colorlinks,linkcolor=blue,anchorcolor=blue,citecolor=blue]{hyperref} 
\hypersetup{
   colorlinks=true, 
   linkcolor=blue,  
   citecolor=blue,  
   filecolor=blue,  
   urlcolor=blue    
} 

\DeclareMathSymbol{\NS}{\mathord}{AMSb}{"4E}

\DeclareSIUnit{\fm}{\femto\meter}

\newcommand{\beq}{\begin{equation}}
\newcommand{\eeq}{\end{equation}}
\newcommand{\beqn}{\begin{eqnarray}}
\newcommand{\eeqn}{\end{eqnarray}}
\newcommand{\bsub}{\begin{subequations}}
\newcommand{\esub}{\end{subequations}}
\newcommand{\bpm}{\begin{pmatrix}}
\newcommand{\epm}{\end{pmatrix}}

\newcommand\identity{1\kern-0.25em\text{l}}

\makeatletter

\newcommand{\Rmnum}[1]{\expandafter\@slowromancap\romannumeral #1@}
\makeatother

\begin{document}

\begin{frontmatter}

\title{Sensitivity of neutrinoless double beta decays from a combined analysis of ground and excited states}
 
\author{C. R. Ding\fnref{1,2}}\ead{dingchr3@mail2.sysu.edu.cn} 
\author{K. Han\fnref{3,4,5}}\ead{ke.han@sjtu.edu.cn}
\author{S. B. Wang\fnref{3,4,5,6}}\ead{shaobo.wang@sjtu.edu.cn}
 \author{J. M. Yao\fnref{1,2}\corref{cor1}}
    \ead{yaojm8@sysu.edu.cn}
    \cortext[cor1]{Corresponding author}
\address[1]{School of Physics and Astronomy, Sun Yat-sen University, Zhuhai 519082, P.R. China}  
  \address[2]{Guangdong Provincial Key Laboratory of Quantum Metrology and Sensing, Sun Yat-Sen University, Zhuhai 519082, China }  
  \address[3]{State Key Laboratory of Dark Matter Physics, Key Laboratory for Particle Astrophysics and Cosmology (MoE), School of Physics and Astronomy, Shanghai Jiao Tong University, Shanghai, 200240, China}  
  \address[4]{Jinping Deep Underground Frontier Science and Dark Matter Key Laboratory of Sichuan Province, Xichang, 615000, China}  
  \address[5]{Shanghai Jiao Tong University Sichuan Research Institute, Chengdu 610213, China}  
  \address[6]{SJTU Paris Elite Institute of Technology, Shanghai Jiao Tong University, Shanghai, 200240, China }

\date{\today}

\begin{abstract} 
Next-generation neutrinoless double-beta ($0\nu\beta\beta$) decay experiments, with projected half-life sensitivities approaching $10^{28}$ years, aim to probe the entire parameter space of the inverted neutrino mass ordering in the light-neutrino-exchange scenario. However, this reach remains uncertain by the substantial model dependence of the nuclear matrix elements (NMEs). In this work, we propose a strategy based on a combined analysis of $0\nu\beta\beta$ decays to both the ground state and the first excited $0^+$ state of the daughter nucleus. We show that such a multi-channel approach can significantly enhance experimental sensitivity, depending on the underlying NME predictions. This method is particularly well suited for large liquid xenon detectors, such as the proposed PandaX-xT and XLZD experiments, which can efficiently identify transitions of \nuclide[136]{Xe} to excited states. Our results highlight the importance of exploiting multiple decay channels in future $0\nu\beta\beta$ searches to maximize their discovery potential.  
\end{abstract}


\end{frontmatter}

\section{Introduction.}
Neutrinoless double-beta ($0\nu\beta\beta$) decay is a hypothetical second-order weak process in which an even-even nucleus transforms into its isobar with two additional protons and two fewer neutrons, accompanied by the emission of only two electrons~\cite{Furry:1939}. Observation of this rare decay would constitute direct evidence of lepton number violation and imply that neutrinos are Majorana particles, possessing a nonzero Majorana mass term~\cite{Schechter:1982PRD}. If mediated predominantly by the exchange of light Majorana neutrinos, the decay rate would allow for the determination of the effective Majorana neutrino mass ($|m_{\beta\beta}|$)~\cite{Agostini:2022RMP}. Although no positive signal has been observed so far, current experiments have set stringent lower bounds on $0\nu\beta\beta$-decay half-lives~\cite{KamLAND-Zen:2022tow,GERDA:2020,CUORE:2021mvw,CUPID:2022puj,EXO-200:2019}. Next-generation ton-scale experiments aim to extend the half-life sensitivity to $\sim$$10^{28}$ years~\cite{Dolinski:2019,Cirigliano:2022oqy,Adams:2022White_Paper}, which would probe the entire parameter space associated with the inverted neutrino mass ordering, depending on the values of the nuclear matrix elements (NMEs). However, current NME calculations carry uncertainties of up to a factor of three or more~\cite{Engel:2017,Yao:2021Review,Cirigliano:2022JPG,Agostini:2022RMP}, presenting a significant obstacle ~\cite{Adams:2022White_Paper,Agostini:2022RMP}. Given the difficulty of directly reducing NME uncertainties, it is of great interest to explore alternative strategies for further enhancing the sensitivity of $0\nu\beta\beta$ searches in ton-scale experiments.
 
Current limits on the effective neutrino mass are typically inferred from the half-life sensitivity of the ground-state-to-ground-state neutrinoless double-beta ($0\nu\beta\beta$-gs) decay channel. An interesting and timely question is whether one can exploit the ground-state-to-excited-state ($0\nu\beta\beta$-ex) decay channel to enhance the sensitivity of next-generation experiments. The ratio of the decay rates for these two  channels are determined by their  $G^{0\nu}_i |M^{0\nu}_i|^2$, where $G^{0\nu}_i$ and $M^{0\nu}_i$  are the phase-space factor (PSF) and  NME of the $i$-th decay  channel, respectively. According to Ref.\cite{Kotila:2012}, the PSFs for the $0\nu\beta\beta$-ex decays in the candidate nuclei are reduced by  factors ranging from approximately 2 to 80. Therefore, provided that the NMEs for the two decay  channels are of comparable magnitude, the observation of the $0\nu\beta\beta$-ex channel could offer additional sensitivity to next-generation $0\nu\beta\beta$ searches.  Indeed, it was found in  \nuclide[150]{Nd}~\cite{Beller:2013PRL} that the NME for $0\nu\beta\beta$-ex decay can even exceed that of the ground-state decay, if the dominant configuration of the excited $0^+$ state more closely resembles the shape of the mother nucleus than the ground state does~\cite{Rodriguez:2010}. 
This possibility is further motivated by the experimental observation of ground-state-to-excited-state two-neutrino double-beta decay ($2\nu\beta\beta$-ex), despite the fact that the corresponding PSF suppression is considerably more severe than in $0\nu\beta\beta$ decay. Indeed, the $2\nu\beta\beta$-ex decay has been observed in \nuclide[100]{Mo}~\cite{Barabash:1995,CUPID-Mo:2023PRC} and \nuclide[150]{Nd}~\cite{Kidd:2014PRC,NEMO-3:2023EPJC,Barabash:2025}, with half-lives roughly two orders of magnitude longer than those of the corresponding ground-state transitions~\cite{NEMO-3:2016qxo,NEMO-3:2020mcq,CUPID-Mo:2023lru}. Recent searches have extended to \nuclide[136]{Xe}\cite{EXO-200:2023pdl,PandaX:2025} and \nuclide[76]{Ge}\cite{Arnquist:2024tsi}. It is worth noting that $0\nu\beta\beta$-ex decay has been explored theoretically in several candidate nuclei~\cite{Tomoda:2000PLB,Simkovic:2001,Simkovic:2001ft,Duerr:2011,Song:2014}. This decay  channel has been proposed either as a way to discriminate among different decay mechanisms~\cite{Simkovic:2001ft} or as a self-consistency test to determine whether an observed signal originates from genuine $0\nu\beta\beta$ decay rather than from other nuclear processes~\cite{Duerr:2011}. 

In this Letter, we carry out realistic simulations of both the $0\nu\beta\beta$-gs and $0\nu\beta\beta$-ex decays of $\nuclide[136]{Xe}$ using the XLZD~\cite{XLZD:2024pdv} detector configurations  and demonstrate that the combined analysis of both decay channels can enhance the sensitivity to $|m_{\beta\beta}|$ without requiring an increase in detector size.  The key advantage arises from the distinctive $0\nu\beta\beta$-ex signature in liquid-xenon time projection chambers (TPCs): the characteristic sequential gamma emissions ($0_2^+ \to 2_1^+ \to 0_1^+$; see Fig.~\ref{fig:cartoon_diagram}(a)) enable both a larger fiducial volume (FV) and more effective background suppression than in the $0\nu\beta\beta$-gs search. 
Consequently, next-generation natural liquid xenon (NNLXe) detectors such as PandaX-xT and XLZD can efficiently identify $0\nu\beta\beta$-ex events and exploit the $\beta\beta+\gamma+\gamma$ signature with excellent spatial and energy resolution, making them uniquely suited for this search.

\section{Simulation of $0\nu\beta\beta$-ex decays in \nuclide[136]{Xe}.} 

\begin{figure}[bt]
\includegraphics[width=0.45\columnwidth]{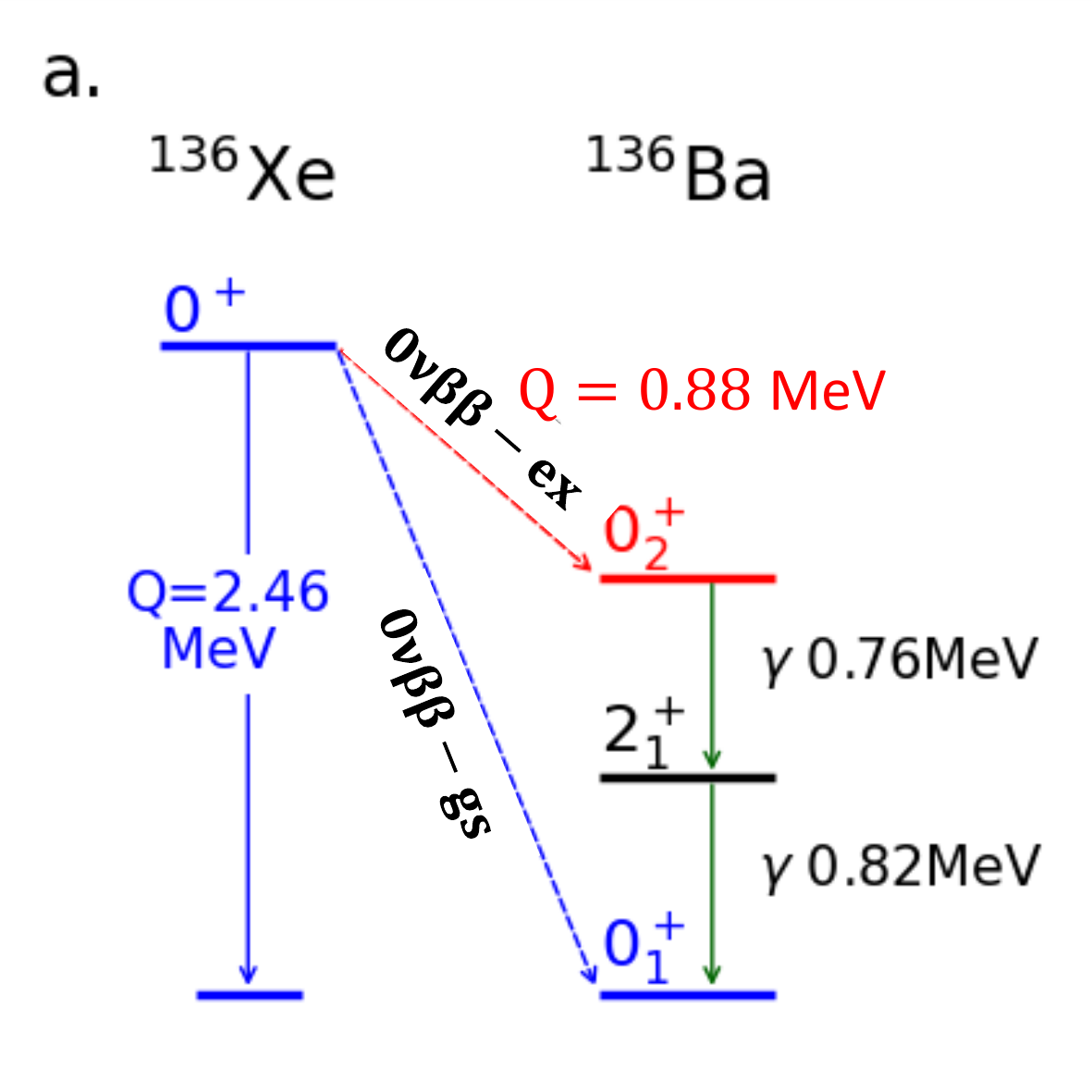} 
\includegraphics[width=0.54\columnwidth]{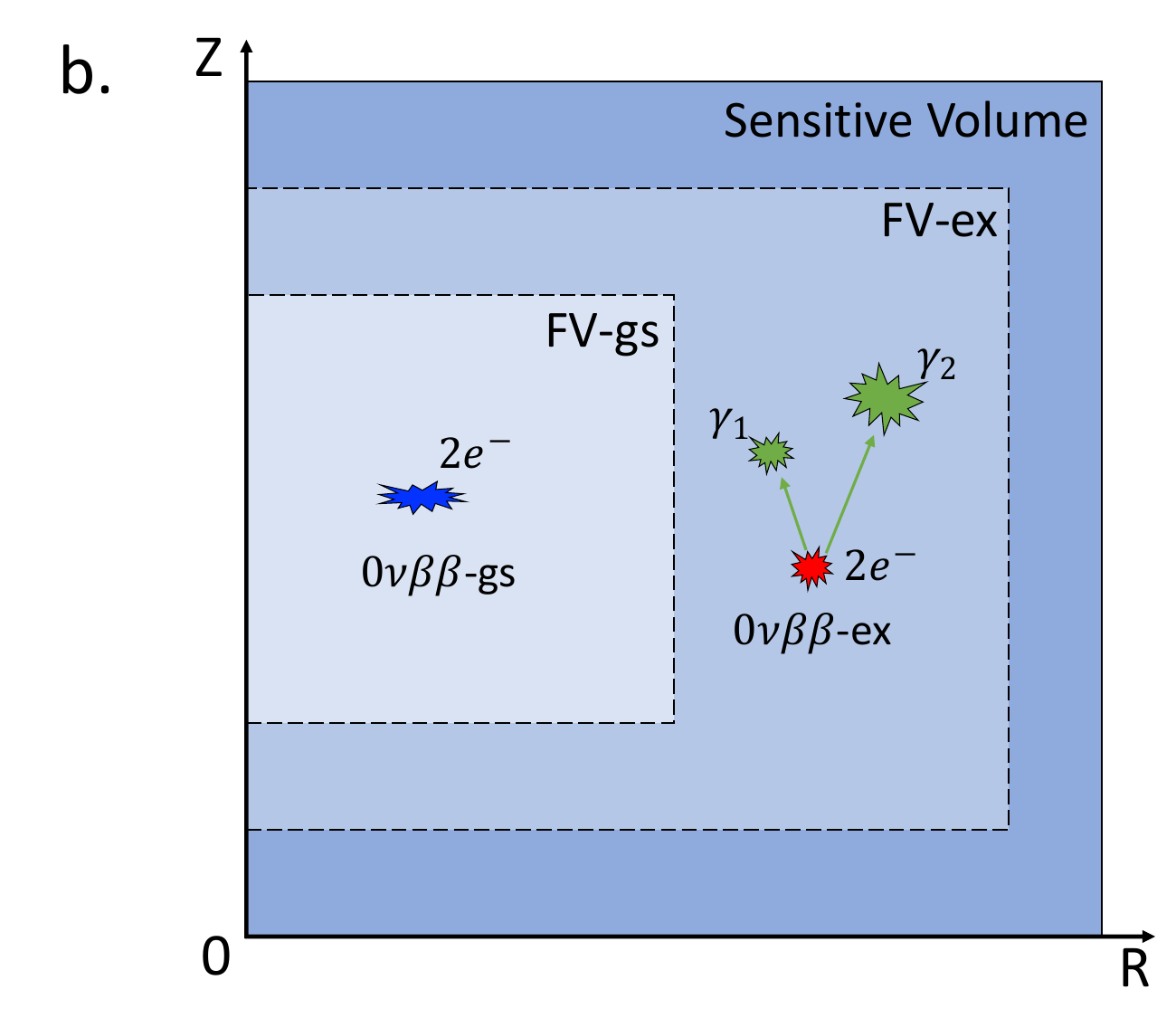} 
     \caption{(Color online) (a) Schematic illustration of the $0\nu\beta\beta$ decay of $\nuclide[136]{Xe}$ to both the ground state ($0^+_1$) and the excited state ($0^+_2$) of $\nuclide[136]{Ba}$, with subsequent gamma emissions ($\gamma_1 = 0.76$ MeV, $\gamma_2 = 0.82$ MeV) following the excited-state transition. (b) Signal signatures in an NNLXe TPC for both $0\nu\beta\beta$-gs and $0\nu\beta\beta$-ex events and illustration of FVs for both decay channels.  The $0\nu\beta\beta$-ex channel produces multi-site events with spatially separated energy depositions from the two electrons and accompanying gammas, while $0\nu\beta\beta$-gs events appear as single-site energy depositions localized near the $Q$-value. 
     Dimensions of FVs are not to scale.}
     \label{fig:cartoon_diagram}
 \end{figure}

The proposed PandaX-xT~\cite{PANDA-X:2024dlo} and XLZD~\cite{XLZD:2024pdv} experiments strategically combine dark matter search and $0\nu\beta\beta$ efforts with 43 and 60 tonnes of natural xenon in the active volume, respectively.
The NNLXe detectors utilize TPC technology and measure both the three-dimensional position and the energy deposition of an event inside the sensitive volume.
The $0\nu\beta\beta$-ex decay of $^{136}$Xe, characterized by a Q-value of $0.88$ MeV and the emission of two de-excitation $\gamma$ rays ($0.76$ MeV and $0.82$ MeV), is illustrated in Fig.~\ref{fig:cartoon_diagram}(a).
As shown in Fig.~\ref{fig:cartoon_diagram}(b), typical $0\nu\beta\beta$-ex decays deposit energy via the continuous double-$\beta$ tracks and $\gamma$s' scattering and/or absorption at different interaction sites. 
The TPC identifies $0\nu\beta\beta$-ex decays as multi-site (MS) events with characteristic energies at each site.
In contrast, $0\nu\beta\beta$-gs events manifest as single-site (SS) events most of the time.

The multi-site nature of $0\nu\beta\beta$-ex decays enables effective background suppression and a substantial increase in the FV available for the search. In $0\nu\beta\beta$ decay experiments, the background rate is a primary factor determining scientific reach. 
For NNLXe detectors, the dominant background originates from the radioactivity of external detector components. 
Owing to xenon's strong self-shielding, this external background exhibits pronounced position dependence. 
To reduce external background contributions and maximize $0\nu\beta\beta$-gs decay search sensitivity, stringent FV cuts (denoted as FV-gs) of NNLXe detectors retain less than 20\% of the xenon target, as in PandaX-xT and XLZD. 
In contrast, the MS signature of $0\nu\beta\beta$-ex decays permits looser FV cuts (denoted as FV-ex), expanding the fiducial xenon mass by roughly a factor of three while lowering the background rate. 
This combination of increased FV and reduced backgrounds leads to a substantial improvement in half-life sensitivity for $0\nu\beta\beta$-ex decay searches.

The numbers of signal events $S_{i}$ and background events $B_{i}$ for both $0\nu\beta\beta$-gs and $0\nu\beta\beta$-ex decays are determined, respectively, by~\cite{GERDA2017}
\begin{equation}
\label{eq:sgnandbkg}
\begin{aligned}
S_{i} &= \ln 2 \cdot \frac{N_A \cdot \epsilon_{i} \cdot \eta_{i}}{m_a}\cdot [T^{0\nu,i}_{1/2}]^{-1},\\
B_{i} &= \eta_{i} \cdot \text{BI}_{i} \cdot \Delta E_{i},
\end{aligned}
\end{equation}
where $i$ labels different decay channels, i.e., the $0\nu\beta\beta$-gs or $0\nu\beta\beta$-ex decay, $N_A$ is Avogadro's number, $\epsilon_{i}$ is the signal efficiency of the $i$-th decay mode, and $\eta_{i} = a M_{i} t_{i}$ is the isotopic exposure [ton·yr], with $a$ the isotopic abundance, $M_{i}$ the fiducial mass [ton], and $t_{i}$ the measurement time. The quantity $m_a$ is the molar mass of the candidate nucleus, $\text{BI}_{i}$ is the background index [cts/(keV·ton·yr)], and $\Delta E_{i}$ is the energy region of interest (ROI) [keV].

To extract the expected sensitivity to $0\nu\beta\beta$, we consider realistic detector response and background expectations based on the published experimental configurations for the XLZD~\cite{XLZD:2024pdv}.
A ``nominal scenario" of the NNLXe detector with an active target of 60~tons is constructed with a Geant4-based simulation framework, BambooMC~\cite{Chen:2021asx}. 
The XLZD detector parameters are adopted for the SS background spectrum, including a ground-state fiducial volume (FV-gs) of 8.2~tons, a background rate of 0.315 counts per year, and an energy resolution of 0.65\% at the $Q$ value ($2.46$~MeV) of the \nuclide[136]{Xe} $0\nu\beta\beta$-gs decay. The detector is assumed to achieve a signal efficiency of 0.76 within a 50-keV ROI centered at the $Q$ value. The $0\nu\beta\beta$-ex signal efficiency and the MS background spectrum are simulated using our detector configuration. Because the dominant \nuclide[238]{U} background peak at 2447~keV lies very close to the decay Q value of 2458~keV and falls within the ROI, we assume that the background is dominated by the \nuclide[238]{U} decay chain. Specifically, the background is modeled by an equivalent external \nuclide[238]{U} activity of 32~mBq placed immediately outside the liquid-xenon sensitive volume, while internal radioactivity is neglected.
Energy depositions in BambooMC are grouped as one site if the distance in the $Z$ direction is less than 5~mm.
For each site, energy is smeared assuming the relative energy resolution is proportional to $1/\sqrt{E[{\rm MeV}]}$ and 0.65\% at $2.46$~MeV.
An event is identified as a $0\nu\beta\beta$-ex signal if the total energy is within the ROI and energy at one site (or the sum of energies at multiple sites) is 0.88~MeV, unless energy at another site is larger than 0.83~MeV.
The straightforward MS-based selection achieves approximately 0.6 signal efficiency, while reducing the background by three orders of magnitude.
Consequently, the $0\nu\beta\beta$-ex decay search benefits from an enlarged FV to 20~tons, in which the background rate is down to 0.126 events per year.

It is noteworthy that the current rudimentary MS selection cuts can be improved for more background suppression power. 
Identification of interaction sites can be improved with three-dimensional information, instead of just the $Z$ direction. 
Clustering of different sites of the Compton scattering and absorption of $\gamma$-rays reconstructs the energy, which provides stringent cuts on $\gamma_1$ and $\gamma_2$ energies.
Machine learning may further exploit the topological signature of $0\nu\beta\beta$-ex events for particle identification. 
We also consider an ideal scenario in which a 100~tons natural liquid xenon (NNLXe) detector with a 60~tons excited-state fiducial volume (FV-ex) is constructed, to illustrate the full potential of the combined $0\nu\beta\beta$-ex and $0\nu\beta\beta$-gs analysis. The background rate in the FV-ex is assumed to be $10^{-4}$ counts per year. The parameters for the $0\nu\beta\beta$-gs channel are taken to be the same as those of XLZD~\cite{XLZD:2024pdv}. For clarity, Table~\ref{tab:para} summarizes the parameters adopted for both the $0\nu\beta\beta$-gs and $0\nu\beta\beta$-ex decays under the nominal and ideal scenarios considered in this work.

\begin{table}[tb]
  \centering 
  \caption{Parameters used in the combined analysis. The values for $0\nu\beta\beta$-gs decay are taken from the XLZD~\cite{XLZD:2024pdv} experiments.}
  \label{tab:para}  
  \setlength{\tabcolsep}{3pt}
  \begin{tabular}{c|ccc}
    \toprule
     &\multirow{2}{*}{$0\nu\beta\beta$-gs}& $0\nu\beta\beta$-ex & $0\nu\beta\beta$-ex \\
     && (nominal) & (ideal) \\
    \midrule
     Isotopic abundance $a$ & $8.86\%$ & $8.86\%$ & $8.86\%$\\
    Efficiency $\epsilon$  & $0.76$ & $0.6$ & $0.6$ \\
    Fiducial volume $M$ [ton] & $8.2$ & $20$ & $60$\\
    Background rate $B$ [cts/yr] & $0.3150$& $0.1260$ & $0.0001$\\
    \bottomrule
  \end{tabular}
\end{table}

\begin{table*}[htbp]
  \centering
  \caption{Comparison of the effective neutrino mass limits (meV) obtained from the $0\nu\beta\beta$ ground-state transition, $|m_{\beta\beta}^{\mathrm{gs}}|$, and from the combined multi-transition analysis, $|m_{\beta\beta}^{\mathrm{comb}}|$, under nominal and ideal scenarios. The values of $|m_{\beta\beta}^{\mathrm{gs}}|$ differ slightly from those reported by XLZD~\cite{XLZD:2024pdv}, owing to different methods for determining the $3\sigma$ sensitivity. The uncertainties of the $|m_{\beta\beta}^{\mathrm{gs/comb}}|$  sensitivities in the MCM method originate from the uncertainties of the NMEs~\cite{Suhonen:2011}, whereas the upper and lower limits of the $|m_{\beta\beta}^{\mathrm{gs/comb}}|$ sensitivities arise from two different treatments for nuclear pairing correlations in the MR-CDFT calculation for the NMEs~\cite{Ding:2023}.
  }
  \label{tab:mbb}
  \setlength{\tabcolsep}{4pt}
  \begin{tabular}{l|ccccccc}
    \toprule
     & RQRPA(RCM) & MCM(Jastrow)& MCM(UCOM)& RQRPA(BEM) & IBM-2 & MR-CDFT & ISM \\
    \midrule
   $| m^{\text{gs}}_{\beta\beta}|$   &$80.1$&$22.4^{+2.4}_{-1.8}$&$16.7^{+1.4}_{-1.2}$&$80.1$&$17.3$&$[10.4, 23.3]$&$29.9$\\
     $| m^{\text{comb}}_{\beta\beta}|$ (nominal)  &$34.6$&$20.4^{+2.4}_{-1.8}$&$16.0^{+1.3}_{-1.1}$&$80.0$&$17.3$&$[10.4, 23.3]$&$29.9$\\
     $|m^{\text{comb}}_{\beta\beta}|$ (ideal) &$8.4$&$7.2^{+1.5}_{-1.0}$&$6.8^{+1.3}_{-0.9}$&$64.2$&$15.2$&$[9.9, 20.4]$&$29.0$\\
    \bottomrule
  \end{tabular}
\end{table*}

\section{Combined analysis of sensitivity to neutrino mass.}
\label{sec:results}
In order to calculate the sensitivity to effective neutrino mass $|m_{\beta\beta}|$, we follow the approach in Ref.~\cite{Pompa:2023jxc} and construct the following $\chi^2$ function:
\begin{equation}
    \begin{aligned}
    \Delta\chi^2&=-2[\ln\mathcal{L}(N|B)-\ln\mathcal{L}(N|N)]\\
    &=2\sum_{i}\bigg[N_{i }\ln\bigg(1+\frac{S_{i}}{B_{i}}\bigg)-S_{i}\bigg],
    \end{aligned}
\end{equation}
where the total events $N_{i}=S_{i}+B_{i}$. The likelihood function is constructed using a Poisson distribution. In the combined analysis, a requirement of $\Delta\chi^2 \geq 9$ defines the region of $|m^{\text{comb}}_{\beta\beta}|$ where a positive $0\nu\beta\beta$ signal can be established at the $3\sigma$ confidence level.

\begin{figure}[bt]
\includegraphics[width=\columnwidth]{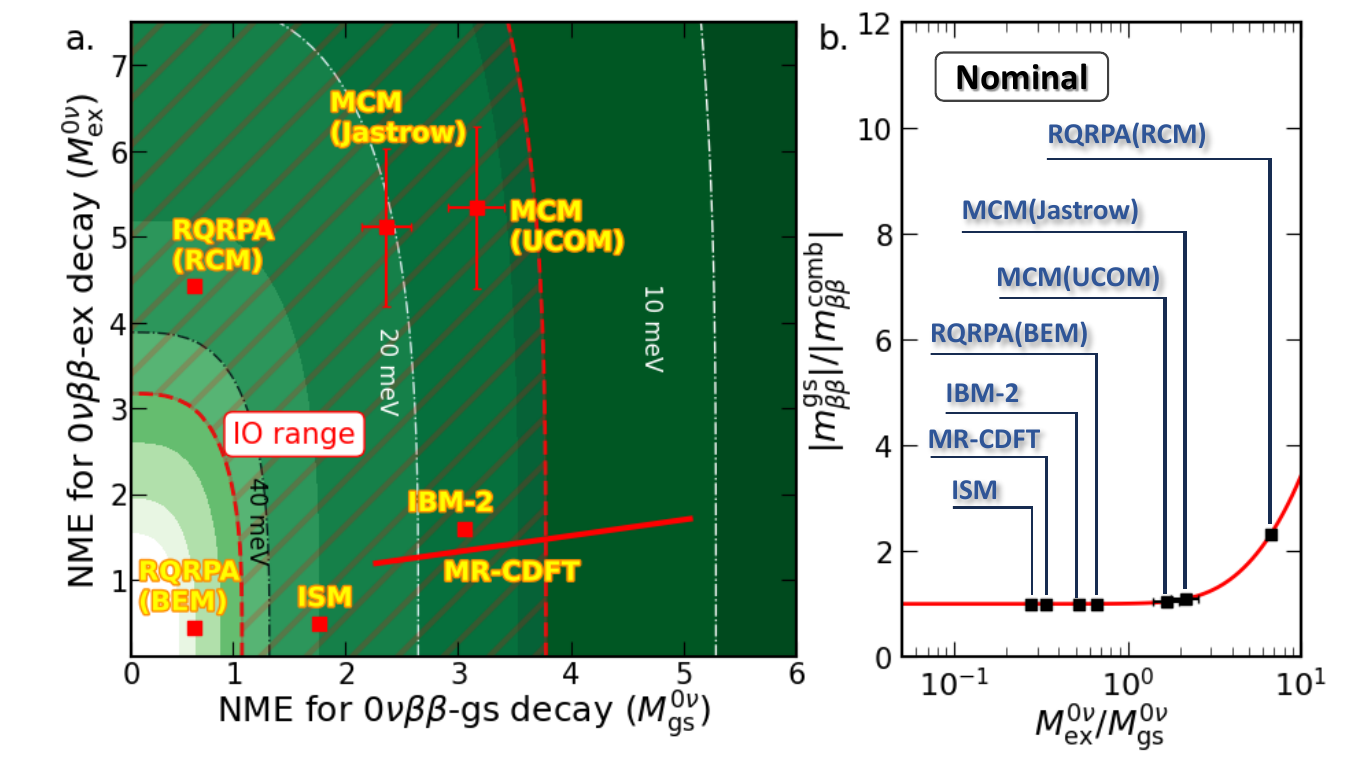} 
\includegraphics[width=\columnwidth]{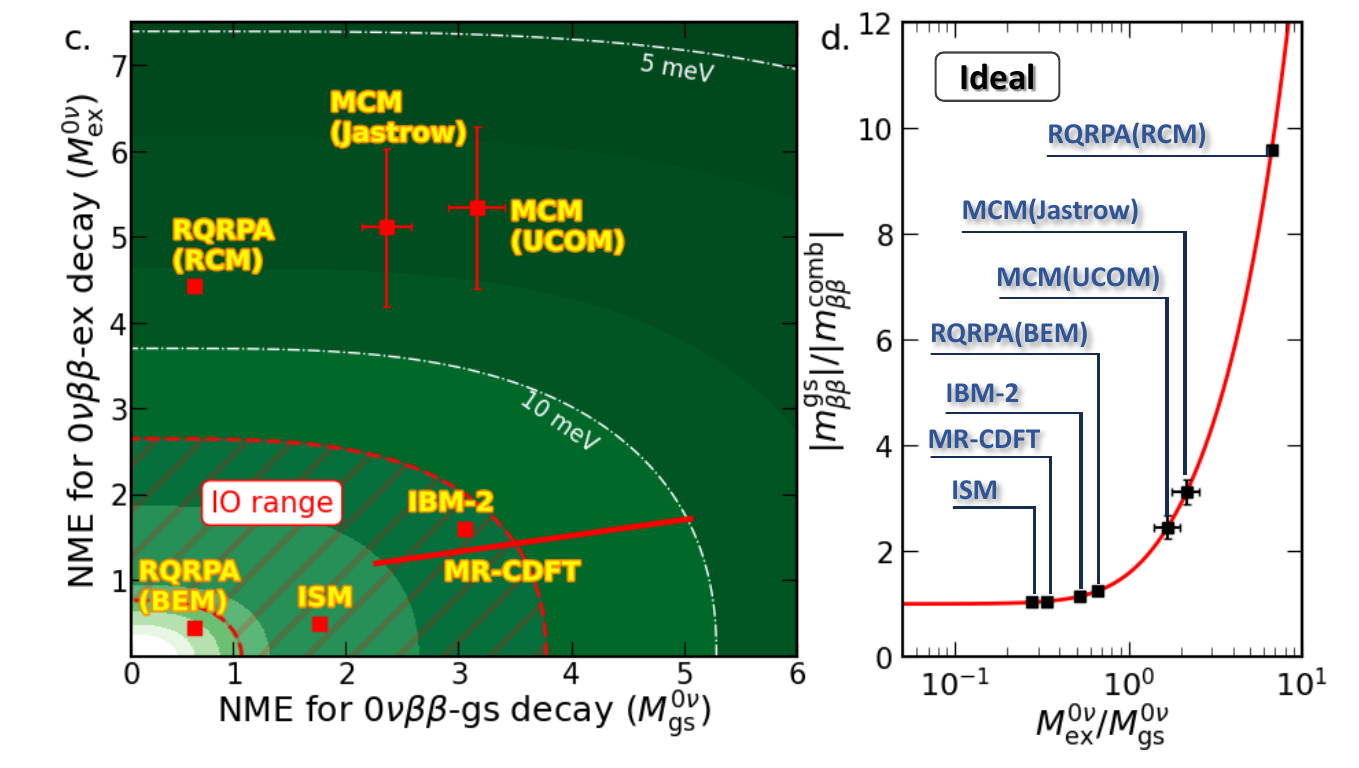} 
     \caption{(Color online) Constraints on the effective neutrino mass obtained from a combined analysis $|m_{\beta\beta}^{\text{comb}}|$ as functions of both $M^{0\nu}_{\text{gs}}$ and $M^{0\nu}_{\text{ex}}$. This analysis incorporates both the nominal scenario (a) and the ideal scenario (c).  
      The red shaded region represents the parameter space for the inverted mass ordering (IO) scenario corresponding to a minimal neutrino mass of zero. The NMEs calculated by different nuclear models are annotated.
     (b) and (d) show the corresponding ratio of constraints on the effective neutrino mass from considering only $0\nu\beta\beta$-gs decay $|m_{\beta\beta}^{\text{gs}}|$ versus the combined analysis $|m_{\beta\beta}^{\text{comb}}|$ as a function of $M^{0\nu}_{\text{ex}}/M^{0\nu}_{\text{gs}}$.}
     \label{fig:mbbSens_XLZD}
 \end{figure}

 The half-life of $0\nu\beta\beta$ decay for each decay channel is given by
\begin{equation}
[T^{0\nu,i}_{1/2}]^{-1} = g_A^4 G^{0\nu}_{i} |M^{0\nu}_{i}|^2 \frac{|m_{\beta\beta}|^2}{m_e^2},
\end{equation}
where $g_A = 1.27$,  the PSF for the $i$-th decay mode, with $G^{0\nu}_{\text{gs}}=14.58\times10^{-15}$yr$^{-1}$ and $G^{0\nu}_{\text{ex}}=0.6127\times10^{-15}$yr$^{-1}$~\cite{Kotila:2012}, and $M^{0\nu}_{i}$ is the corresponding NME. For \nuclide[136]Xe, the NMEs for both decay channels  have been predicted by several nuclear models. These include the renormalized quasiparticle random-phase approximation (RQRPA)~\cite{Simkovic:2001}, supplemented with two different treatments of the excited $0^+$ state of the daughter nucleus—the recoupling method (RCM)~\cite{Griffiths:1992} and the boson expansion method (BEM)~\cite{Raduta:1991,Raduta:1996}; the multiple-commutator model (MCM)~\cite{Suhonen:2011,Hyvarinen:2016udc}, in which nucleon–nucleon short-range correlations are incorporated using either the UCOM correlator~\cite{Kortelainen:2007PLB} or the Jastrow correlator~\cite{Miller:1976AP}; the interacting boson model (IBM-2)~\cite{Barea:2015}; the interacting shell model (ISM)~\cite{Menendez:2009}; and the multi-reference covariant density functional theory (MR-CDFT)~\cite{Ding:2023}. Within the MR-CDFT framework, two sets of NMEs were obtained using different treatments of nuclear pairing correlations~\cite{Ding:2023}. In the following analysis, both sets are used in the combined sensitivity study of $|m_{\beta\beta}|$, and the resulting sensitivities are taken to define the lower and upper bounds of the MR-CDFT predictions.
We note that some of these NME calculations rely on crude approximations and may change when additional correlations are treated more carefully. Moreover, it remains difficult at present to determine which calculation is the most reliable. Given this situation, and in order to draw a meaningful conclusion, we include all of the available NMEs in our analysis and interpret the spread in the predicted sensitivity enhancement as an estimate of the model-dependent uncertainty. We find that the NMEs for the $0\nu\beta\beta$-gs decay ($M^{0\nu}_{\text{gs}}$) range from 0.66 to 5.06, while those for the $0\nu\beta\beta$-ex decay ($M^{0\nu}_{\text{ex}}$) span from 0.49 to 6.28. For both decay channels, the NMEs vary by up to an order of magnitude. See Fig. 1 of the Supplemental Material for the comparison of these NMEs. The PSFs are taken from Ref.\cite{Mirea:2015nsl}.

  \begin{figure*}[bt]
  \includegraphics[width=2\columnwidth]{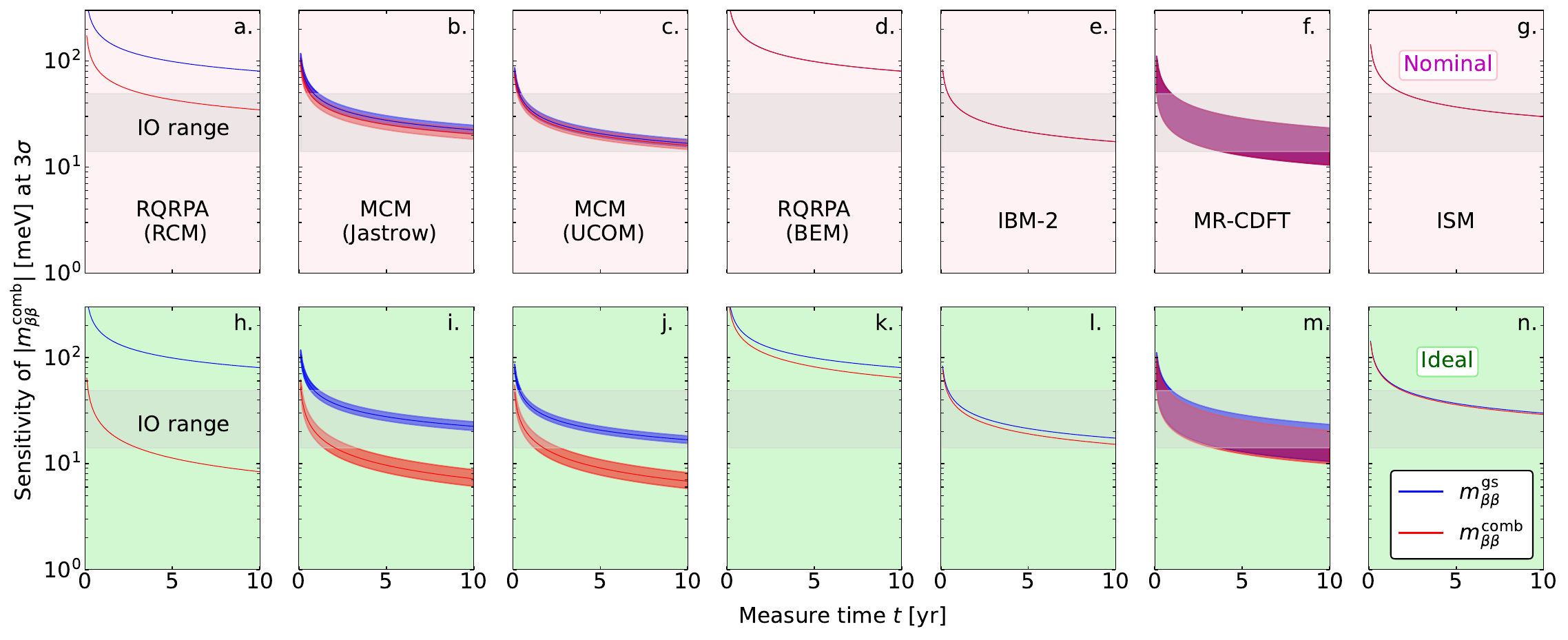} 
     \caption{(Color online) The projected $3\sigma$ sensitivity limits on the effective neutrino mass $|m_{\beta\beta}|$ as a function of measurement time, comparing results from the ground-state-only channel ($m^{\text{gs}}_{\beta\beta}$, blue) and the combined analysis including the excited state ($m^{\text{comb}}_{\beta\beta}$, red), across different nuclear models, for a fixed background count. The top panels (a–g) show results under the nominal scenario, while the bottom panels (h–n) show results under the ideal scenario; shaded bands represent the range of predictions due to model uncertainties, and the IO region is marked for reference. The difference between the $m_{\beta\beta}^{\rm gs}$ and $m_{\beta\beta}^{\rm comb}$ in panels (d-g) is too small to be seen evidently, which is consistent with the Fig.~\ref{fig:mbbSens_XLZD}(b).
     }
     \label{fig:Sensitivity}
 \end{figure*}

Assuming a 10-year data acquisition period, we evaluate the sensitivity to $|m_{\beta\beta}|$ for nominal and ideal scenarios, as summarized in Table~\ref{tab:mbb} and Fig.~\ref{fig:mbbSens_XLZD}. 
The results indicate that the upper limits of the NMEs obtained with MR-CDFT yield the most stringent constraints on $|m_{\beta\beta}|$, both when considering only the $0\nu\beta\beta$ decay to the ground state and when performing the combined analysis in the nominal scenario. In both cases, the projected lower sensitivities to $|m_{\beta\beta}|$ reach $\sim 10$ meV after 10 years of operation, falling below the lower bound expected for the inverted neutrino-mass ordering.
However, under the ideal combined analysis scenario, the analysis with NMEs of MCM (UCOM) establishes the most stringent constraint of $6.8^{+1.3}_{-0.9}$~meV, where the uncertainties originate from those of the NMEs.
Fig.~\ref{fig:mbbSens_XLZD}(b(d)) shows how the enhancement factor varies with the NME ratio for the $0\nu\beta\beta$-ex and $0\nu\beta\beta$-gs transitions for the nominal (ideal) scenario. 
The enhancement in sensitivity increases with the NME ratio, and the exact improvement depends on the experimental parameters.  
The combined analysis improves the sensitivity of $|m_{\beta\beta}|$ from $80.1$ meV to $34.6$ meV, bringing it well within the inverted ordering (IO) region,  with NMEs from the RQRPA (RCM) approach.
A meaningful improvement is achieved when $M^{0\nu}_{\text{ex}} \geq M^{0\nu}_{\text{gs}}$ for the nominal scenario, while the improvement is much more significant for the ideal scenario. For the NMEs calculated with RQRPA (RCM), MCM (Jastrow), and MCM (UCOM), the sensitivity improvement can be as significant as 2 to 10. Based on these NMEs and the assumption that $0\nu\beta\beta$ decay is dominated by the light-neutrino-exchange mechanism,  the combined analysis could bring the $|m_{\beta\beta}|$ sensitivity to below 10~meV, enabling $0\nu\beta\beta$ experiments to fully cover the IO region within a 10-year measurement period, and potentially allowing for the determination of the neutrino mass ordering.

Figure~\ref{fig:Sensitivity} presents the $3\sigma$ sensitivity to the effective neutrino mass as a function of detection time, using NMEs from various models.
The combined analysis enhances sensitivity for all models, with the improvement most pronounced under the ideal scenario. 
Under the nominal scenario, the IO region is partially covered, whereas under the ideal scenario, it is fully covered within a few years.  These results suggest that, within current or near-future capabilities, the combined multi-transition analysis of $\nuclide[136]{Xe}$ could significantly accelerate experimental access to the IO regime, potentially reaching or surpassing IO sensitivity within the next decade. We further examine the impact of the $0\nu\beta\beta$-ex background rate on the combined analysis in the Supplemental Material.
The improvement of the combined analysis can be obtained beyond the nominal and ideal scenarios.

\section{Summary and outlook.}
Next-generation ton-scale $0\nu\beta\beta$ experiments are under development with the goal of probing the lower boundary of the inverted-mass-ordering region, 
but the large uncertainties in NMEs remain a major limiting factor. Under these conditions, any improvement in sensitivity becomes particularly valuable. In this Letter, we have proposed a strategy to enhance the sensitivity  to the effective Majorana mass in such experiments by performing a combined analysis of ground-state and excited-state $0\nu\beta\beta$ decays in a single experiment. Large natural-xenon time projection chambers are particularly well suited for detecting excited-state decays with high efficiency. Using realistic experimental parameters for \nuclide[136]{Xe}, and focusing on the PandaX-xT and XLZD configurations, we estimate reference sensitivities based on illustrative background simulations. While the precise enhancement depends on the choice of NMEs and specific detector settings, the combined analysis has been shown to improve the sensitivity to the effective neutrino mass by more than a factor of two in a nominal scenario and by up to an order of magnitude under ideal conditions. Such an improvement enhances the prospect that next-generation experiments could fully cover the effective neutrino-mass parameter space associated with the inverted ordering. 

Although this work focuses on the combined analysis of \nuclide[136]{Xe}, other candidate nuclei also show a strong potential for sensitivity enhancement. 
The PSF indicates that \nuclide[100]{Mo} and \nuclide[150]{Nd} are particularly promising. 
Unlike \nuclide[136]{Xe}, where the combined analysis significantly improves sensitivity only when $M^{0\nu}_{\text{ex}}$ and $M^{0\nu}_{\text{gs}}$ are similar, \nuclide[150]{Nd} (\nuclide[100]{Mo}) can achieve comparable enhancements at $M^{0\nu}_{\text{ex}}/M^{0\nu}_{\text{gs}} \approx 1/3\,(1/2)$, assuming experimental parameters comparable to those of \nuclide[136]{Xe}. 
This is especially true for \nuclide[150]{Nd}, whose ground-state shape closely resembles that of the excited state in \nuclide[150]{Sm}, potentially leading to a larger NME for decay into the excited $0^+$ state~\cite{Beller:2013PRL}. Moreover, the present work does not include transitions to the $2^+$ state. Previous studies have shown that the PSF for the $0\nu\beta\beta(2^+)$ decay can be comparable to that of the ground-state transition~\cite{Mirea:2015nsl,Tomoda:1988}, suggesting that the $0\nu\beta\beta(2^+)$ channel may contribute non-trivially~\cite{Tomoda:1988,Fang:2023}. This decay channel could be incorporated into future combined analyses.

Finally, we emphasize that the effectiveness of the combined analysis depends strongly on the values of the NMEs, which currently lack systematic uncertainty quantification and exhibit significant discrepancies among different nuclear models. A proper uncertainty assessment could either reduce or increase the improvement factor obtained from the combined analysis, and this remains an open question. Nevertheless, our results demonstrate that a combined analysis of multiple transitions can enhance the sensitivity of next-generation experiments to the effective neutrino mass without requiring a larger detector. A more definitive assessment will become possible as theoretical developments yield more precise NMEs for both ground- and excited-state transitions. This situation further underscores the importance of achieving theoretically accurate and systematically quantified NME calculations for  $0\nu\beta\beta$ decay.

\section*{Acknowledgments} 
We thank T. Li for the help with the simulation. We also thank A. Belley, J. Engel, D. L. Fang, J. Holt, G. Li, Y. K. Wang, and J. Y. Zhu for valuable discussions. This work was supported in part by the Ministry of Science and Technology of China (No. 2023YFA1606202), the National Natural Science Foundation of China (Grant Nos. 125B2108, 12375119 and 12141501), the Guangdong Basic and Applied Basic Research Foundation (2023A1515010936), and the Natural Science Foundation of Shanghai (No. 24ZR1437100). 

 
%

\end{document}